This manuscript is accepted by **MNRAS**

# The future of low-mass condensations in a core of molecular cloud

Mohsen Nejad-Asghar

*Department of Physics, University of Mazandaran, Babolsar, Iran*

*Research Institute for Astronomy and Astrophysics of Maragha, Maragha, Iran*

*Department of Physics, Damghan University of Basic sciences, Damghan, Iran*

## ABSTRACT

Two scenarios have been proposed for evolution of star forming cores: gravitational fragmentation of larger structures and coalescence of smaller entities which are formed from some instabilities. Here, we turn our attention to the latter idea to investigate the evolution of observed low-mass condensations (LMCs) in the cores of molecular clouds. For this purpose, we implement the evolution of the observed LMCs of Taurus molecular cloud 1 (TMC-1). The core is modeled as a contracting cylinder with randomly spawned condensations in the middle region around its axis. For advancing bodies in their trajectories, we represent the acceleration of a particular LMC in terms of a fourth-order polynomial using the predictor-corrector scheme. Whenever two LMCs collide, they are assumed to be merged in one large condensation containing all the masses of the two progenitors. Implementations of many computer experiments with a wide variety of the free parameters show that the LMCs merge to form star-forming regions in the core. The results show that the total mechanical energy of the core increases by time, and its rate of increasing decreases by facilitating the merger. Finally, the mass spectrum index and goodness-of-fit are determined with 50% error in the number of mass points. The results show that the goodness-of-fit will be refined at the end of simulations, and the mass spectrum index inclines to the observed values for the moderate mass objects. The simulations show that the TMC-1 turns about 40% of its mass into cluster of dynamically unstable protostellar cores. In general, we suggest that the future of LMCs in a core of molecular cloud is merger to convert about half of its initial masses into a cluster of gravitationally unstable protostellar cores.

## 1. Introduction

Star formation appears to occur exclusively within the molecular phases of the interstellar medium (see, e.g., Klessen et al. 2009). Molecular clouds exhibit extremely complex hierarchical structure, with column densities that vary by many orders of magnitude. The smallest divisions are nominated as protostellar cores, with masses of a few solar mass or less and sizes of few tenth parsecs or less (e.g., Stahler and Palla 2004). The advent of sensitive tools and advanced observational methods in the last decade has made it possible to perform high-resolution surveys to interior of the individual cores (e.g., Tafalla 2008). In this way, many of embedded condensed objects within each core have been revealed that we call them low-mass condensations (LMCs). Although a lot of these have recently found by the Spitzer c2d project (e.g., Lee et al. 2009), but we document this research on the pioneer report of Peng et al. (1998, hereafter PLVKL) who presented observational data on the small-scale structure of the Taurus Molecular Cloud 1 (TMC-1) in the regime of $0.02 - 0.04$pc and $0.04 - 0.06 M_{\odot}$. This preference is because of many small groups and associations of star formation in the Taurus molecular cloud that provides it as an important test-place for theories and models (Hartman 2002).

The formation mechanisms of LMCs are not well understood. In fact, two schemes have been proposed: gravitational fragmentation of larger structures and merger of smaller fragments. Preserving the latter mechanism requires some suitable theoretical models for the formation of the smaller entities in the cloud. One model that may be regarded is the effect of turbulence, although estimations of the level of turbulence in the molecular cores give subsonic values (e.g., Myers 1998, Lee et al. 2004, André et al. 2007). As another idea, van Loo et al. (2007) implemented a two-dimensional simulation with the idea that the mechanism causing substructures is the same as the mechanism causing core formation (i.e., hydromagnetic waves). The appeared substructures in their simulation are accurately conformed to the observed substructures of core D of TMC-1. Although, the dynamical models like turbulence or hydromagnetic waves can be considered as sources of substructures in dense cores, but we must not neglect some non-dynamical processes and instabilities such as thermal instability (e.g., Fukue and Kamaya 2007, Nejad-Asghar 2007, Nejad-Asghar and Molteni 2008, Banerjee et al. 2009, Nejad-Asghar and Soltani 2009). In any case, formation of a system of LMCs in each protostellar core is a controversial issue. In this research, we have nothing to do with the formation issue of LMCs, rather we attempt to model their future. Since, our aim is to present a realistic model for future evolution of LMCs, we use the data of PLVKL for LMCs in the core D of TMC-1. The results may approximately be applicable to all cores with a system of LMCs.

The high-resolution investigation in the southeastern part of the TMC-1, which was done by PLVKL, covers an elongated $8' \times 8'$ area in the sky, or 0.32pc at the distance of 140pc. The emission in this region originates in three narrow components centered on Local Standard of Rest (LSR) velocities of $\sim 5.7$, $\sim 5.9$, and $\sim 6.1\mathrm{km.s^{-1}}$. These components each represent a separate cylindrical feature elongated along the ridge of TMC-1. Among the three velocity components, PLVKL identified a total of 45 LMCs with masses in the range of $0.04 - 0.06 M_{\odot}$ and sizes of $0.02 - 0.04$pc. PLVKL suggested that thermal and turbulent pressures of the ambient gas of the core may contribute to keep these LMCs pressure bond so that the dispersion time of LMCs is comparable to the molecular cloud's lifetime $\sim 10^6$yr (e.g., Tan et al. 2006). Typical inflow velocities in the protostellar cores are of order $0.05 - 0.1\mathrm{km.s^{-1}}$ (Lee et al. 1999), which is comparable to the results of PLVKL for the random motion of order $0.05\mathrm{km.s^{-1}}$ among the LMCs. We suggest the gravitational force between LMCs can influence the future of them. On the other words, over the lifetime of LMCs, they are more likely to be subject to merge because of subsonic collisions with each other rather than to travel intact along the core. Investigating this future is the task of this research.

In fact, modeling future evolution of LMCs in protostellar cores adequately requires the accurate and simultaneous treatment of many different physical processes. Here, we try to overcome the computing limitations by using an adapted N-body method, in which the magnetic field, shock waves, heating and cooling processes are ignored. The effect of ambient gas on motion of each LMC is assumed to be only a drag force as outlined by Ostriker (1999). Whenever two LMCs approach each other, the result of interaction is determined according to prescribed rules, which have been numerically examined by many authors (e.g., Lattanzio and Henriksen 1988, Gittins et al. 2003, Kitsionas and Whitworth 2007). The simulations, in general, show that in subsonic collisions with small impact parameter, the outcome is always a merger. Burkert and Alves (2009) have recently used this idea to simulate the future of the two LMCs in the protostellar core Barnard 68. Their simulation shows that merger of these two sub-cores will lead to gravitational instability. Thus, merger of LMCs might in general play an important role in triggering star formation and shaping the molecular core mass distribution.

It seems that using a suitable gravitational N-body method for evolution of LMCs in the core D of TMC-1, and using a parametric method for considering the merger of them, can give us some acceptable progress for future of the observed LMCs in the molecular cores. For this purpose, setup of computer experiments with initial positions and velocities of LMCs in the core D of TMC-1 are introduced in section 2. Advancing the LMCs in their trajectories, considering of merger, and the results of simulations are presented in section 3. Finally, section 4 devotes to a summary and conclusion.

## 2. Setup of computer experiments

We choose the units of length and time as $u_l = 3\times10^{14}$m $= 0.01$pc and $u_t = 1.5\times10^4$yr, respectively, therefore the unit of velocity is $u_v = 0.67$km.s$^{-1}$. The mass unit is chosen as $u_m = 2\times10^{30}$kg $= 1M_\odot$ so that the gravitational constant is $G = 1$. Each LMC in the core D of TMC-1 is assumed to be a sphere with average radius $R = 3.39\times10^{-2}(b_{maj}b_{min})^{1/2}u_l$ where a distance of 140pc is used, and $b_{maj}$ and $b_{min}$ are major and minor axes in arc-seconds, respectively, which are listed in Table 1 of PLVKL. The radii of the 45 LMCs and their masses at the present time, $t = 0$, are shown in Fig. 1. If we assume that the LMCs are pressure confined, self-gravitating isothermal gas spheres in hydrostatic equilibrium, we can approximately adopt the relation $R \approx 18.7M10^{-M}$ between the radius and mass of them in units of $u_l$ and $u_m$, respectively (see Appendix A).

The region that analyzed by PLVKL is an $8'\times8'$ area of the TMC-1 which is equivalent to 0.32pc for the cloud at distance of 140pc. The distribution of integrated intensity of the three components b (blue), m (middle), and r (red) are mapped in Figure 2 of PLVKL. Overall, maps show an elongated appearance, which may represent a cylindrical structure for the core. In this way, we consider a parent core in a cylindrical geometry with length $45u_l$ and diameter $15u_l$. The LMCs of blue (b) and middle (m) components are distributed along the length of cylinder, while the LMCs of red (r) component are approximately placed in three quarter of the length. We divide the realm of each LMC by some subdivisions along the length of the parent core, and then its center is randomly placed into these subdivisions. The azimuth angle for position of each LMC is randomly determined between 0 to $2\pi$, and its axial distance is randomly chosen between $f_{in}R_0$ and $f_{out}R_0$ where $R_0 = 7.5u_l$ is the radius of cylindrical core. According to the best matches to the depicted maps of Figure 2 of PLVKL, the fractions $f_{in}$ and $f_{out}$ are obtained approximately equal to 0.4 and 0.7, respectively (see Fig. 2).

As shown in Table 1 and 2 of PLVKL, the speed of LMCs have a small dispersion about the mean with a standard deviation of 0.03km.s$^{-1}$ $= 0.045u_v$. On the other hand, Lee et al. (1999) show that the typical inflow velocities in the molecular cores are of order $0.05-0.1$km.s$^{-1}$. Thus, here we assume the speed of each LMC, $v$, to be randomly generated in the range of $0.03-0.1$km.s$^{-1}$. The velocity vector has two components: a radial part $v_r$ toward the center of the core, and a transverse component $v_\perp$. As depicted in Fig. 3, the velocity in the $x'y'z'$-coordinate is given by

$$\mathbf{v}' = \hat{i}'v_\perp\cos\varphi' + \hat{j}'v_\perp\sin\varphi' + \hat{k}'v_r, \qquad (1)$$

where $\varphi'$ is the azimuth angle.

We introduce two free parameters for velocity vector: (1) the transverse velocity param-eter $\alpha$ so that $v_\perp = \alpha v$ and $v_r = \sqrt{1-\alpha^2}v$, and (2) the inclination parameter $\beta$ so that the azimuth angle $\varphi'$ is randomly produced between $\pm\beta\pi/2$. In this way, the velocity vector $\mathbf{v}'$ for each LMC is determined whenever the values of these two free parameters are appointed. Since, the coordinate $x'y'z'$ is obtained via two successive rotations of $xyz$,

$$R_{y'}(\theta+\pi)R_z(\varphi) = \begin{pmatrix} -\cos\theta & 0 & \sin\theta \\ 0 & 1 & 0 \\ -\sin\theta & 0 & -\cos\theta \end{pmatrix} \begin{pmatrix} \cos\varphi & \sin\varphi & 0 \\ -\sin\varphi & \cos\varphi & 0 \\ 0 & 0 & 1 \end{pmatrix}, \tag{2}$$

the velocity vector $\mathbf{v}$ in $xyz$-coordinate can be determined by

$$\mathbf{v} = \left[R_{y'}(\theta+\pi)R_z(\varphi)\right]^{-1}\mathbf{v}'. \tag{3}$$

Each LMC within the aggregate of bodies in the parent core experiences an acceleration that arises from two sources: (1) the gravitational attractions of other LMCs,

$$\mathbf{a}^{(i)}_{grav} = -\sum_{j=1,j\neq i}^{N} \frac{GM_j(\mathbf{r}_i - \mathbf{r}_j)}{|\mathbf{r}_i - \mathbf{r}_j|^3}, \tag{4}$$

and (2) a drag force from ambient gas,

$$\mathbf{a}^{(i)}_{drag} = -\frac{4\pi}{3}G^2\frac{\rho_o}{c_s^3}M_i\mathbf{v}_i, \tag{5}$$

where $\rho_0$ and $c_s$ are the mass density and sound speed of the ambient gas in the parent core, respectively (Ostriker 1999).

The description of the problem is now completed by specified accelerations for all the LMCs. Thus, the position of the bodies can be advanced by an accurate N-body method. Whenever the impact parameter of two colliding LMCs, with radii $R_i$ and $R_j$, is less than $(R_i + R_j)/\gamma$, they are assumed to be merged in a one larger condensation with radius $\approx 18.7M10^{-M}$ (see Appendix A), containing all the mass of the two progenitors ($M = M_i + M_j$). The collision is assumed to be perfectly inelastic so that the position and velocity after merge are given by

$$\mathbf{r} = \frac{M_i\mathbf{r}_i + M_j\mathbf{r}_j}{M_i + M_j}, \tag{6}$$

and

$$\mathbf{v} = \frac{M_i\mathbf{v}_i + M_j\mathbf{v}_j}{M_i + M_j}, \tag{7}$$

respectively. Note that the properties of the models in each computer experiment are completely controlled by three key parameters $\alpha$, $\beta$, and $\gamma$.

## 3. Results of simulations

Determining the new positions of the LMCs at a somewhat advanced time is deceptively simple in appearance, but a slight carelessness can lead to a complete change of long-term behavior. Here, we use the method of Aarseth (2003) for advancing LMCs in their trajectories. This method is a version of a predictor-corrector scheme, which pivots on representing the acceleration of a particular body of index $i$ at time $t$ in terms of a fourth-order polynomial based on knowledge of the acceleration at four previous times in the past, $t_0$, $t_1$, $t_2$, and $t_3$, with $t_0$ being the most recent,

$$\mathbf{a}(t) = (((\mathbf{D}_4(t-t_3) + \mathbf{D}_3)(t-t_2) + \mathbf{D}_2)(t-t_1) + \mathbf{D}_1)(t-t_0) + \mathbf{a}_0, \tag{8}$$

where $\mathbf{a}_0 = \mathbf{a}_{grav} + \mathbf{a}_{drag}$ is the acceleration at time $t_0$. Using compact notation, the first three divided differences are defined by

$$\mathbf{D}_k[t_0, t_k] = \frac{\mathbf{D}_{k-1}[t_0, t_{k-1}] - \mathbf{D}_{k-1}[t_1, t_k]}{t_0 - t_k}, \quad (k = 1, 2, 3), \tag{9}$$

where $\mathbf{D}_0 = \mathbf{a}$ and square brackets refer to the appropriate time intervals. The term $\mathbf{D}_4$ is defined similarly as

$$\mathbf{D}_4 = \frac{\mathbf{D}_3[t, t_2] - \mathbf{D}_3[t_0, t_3]}{t - t_3}, \tag{10}$$

which the predictor corrector method is used to obtain quantities at the advanced time $t$ that information is not initially available.

When the acceleration of each LMC is calculated from (8), one can immediately integrate with respect to time obtaining its velocity and then again to obtain the position. The key idea underlying to do is determining a suitable time-step. A simple method for estimating the time-step for each LMC involves the distance and the relative velocity with respect to its nearest neighbor. Here, we adopt the more sophisticated relation

$$\triangle t_i = \eta \left( \frac{|\mathbf{a}||\mathbf{a}^{(2)}| + |\mathbf{a}^{(1)}|^2}{|\mathbf{a}^{(1)}||\mathbf{a}^{(3)}| + |\mathbf{a}^{(2)}|^2} \right)^{1/2}, \tag{11}$$

where $\eta$ is a dimensionless accuracy parameter ($\approx 0.01$ in our computer experiments), and the superscript in parentheses is the order of the derivative (i.e., $\mathbf{a}^{(1)} = d\mathbf{a}/dt$, and so on).

The properties of models in the computer experiments are completely controlled by three key parameters: transverse velocity component $0 < \alpha < 1$, axial inclination of the velocity vector $0 \leq \beta \leq 2$, and merger parameter $\gamma \sim 4$. Simulations with different values of these free parameters show that in any case, the LMCs merge to form more massive larger bodies. The results, as expected, show that decreasing of the three parameters $\alpha$, $\beta$, and $\gamma$ leads to facilitating the merger of LMCs and formation of star-forming regions in the parent core. With typical values of $\alpha = 0.1$, $\beta = 1.0$, and $\gamma = 4.0$, the number of LMCs, $N$, and the mean mass, $\sum M_i/N$, at different simulation time are shown in Fig. 4.

In a perfectly inelastic collision between two LMCs which leads to merger, the kinetic energy dissipates and the two-body potential energy approaches to zero. Thus, in each tow-body merger, the total mechanical energy increases. This fact is shown in Fig. 5 for three values of the merger parameter $\gamma$, in which the parameters $\alpha$ and $\beta$ are chosen equal to 0.1 and 1.0, respectively. As can be seen in this figure, the total mechanical energy of a system of LMCs, which are initially gravitationally bond, increases by time until it approaches to zero at a critical time $t_c$. Afterwards, the system will be gravitationally unbound. Fortunately, this critical time is less than the lifetime of LMCs so that the results of our computer experiments, in this simplest implementation, are reliably fruitful at $t \leq t_c$. The rate of increasing of the total mechanical energy of a system of LMCs, in consequence of merger, decreases by decreasing of the parameters $\alpha$, $\beta$, and $\gamma$. On the other words, the critical time $t_c$ increases by facilitating the merger of LMCs in the computer experiments. The computer experiments show that the effect of the parameters $\alpha$ and $\beta$ on the critical time $t_c$ is negligible, although the results indicate that decreasing of the transverse parameter $\alpha$ and inclination parameter $\beta$ can facilitate the merger of LMCs and leads to little increasing of critical time $t_c$.

The mass spectrum of clumps in giant molecular clouds appear to be well described by a power law,

$$\frac{dN}{dM} \propto M^{a-1}, \tag{12}$$

with the index being in the range $-1.8 < a - 1 < -1.3$ (e.g., Kramer et al. 1998). When considering the protostellar cores, the mass function is well described by a double power law fit, following approximately $a - 1 \approx -2.5$ above $\sim 0.5 M_\odot$ and $a - 1 \approx -1.5$ below (Motte et al. 1998, Testi and Sargent 1998, Johnstone et al. 2001, Johnstone et al. 2006, Alves et al. 2007). There can also be found a power law mass function in the low-regime of brown dwarfs (Thies and Kroupa 2008). Overall, the canonical mass function with confidence found so far is a power law function with universality hypothesis that it constitutes the parent distribution of all stellar populations (Kroupa 2008).

In the low-mass regime of LMCs, PLVKL excluded the lowest and highest mass points to study the mass distribution of these small entities by using a power law fit, $\log N = a \log M + const.$, with $a \approx -0.45$. They indicated that the corresponding mass spectrum index $a - 1 = -1.45$ is in reasonable agreement with those obtained by observational data analysis of stellar populations. Here, we assume that there is approximately 50% error in the number of LMCs ($\triangle(\log N) \approx |\log(1.5)|$), and the mass spectrum index $a-1$ is evaluated in the simulation time. The results of this linear fit and its goodness-of-fit (Press et al. 1992)

$$q = Q\left(\frac{N-2}{2}, \frac{\chi^2}{2}\right) \tag{13}$$

where $Q$ is the incomplete gamma function and $\chi^2$ is chi-square, are given in Fig. 6. This figure shows that the goodness-of-fit will be refined at the end of simulation, and the mass spectrum index inclines to the expected values for the moderate mass populations. Fig. 7 shows the histogram of the mass distribution of LMCs at present time, and at the simulation time $t \approx 1.5 \times 10^5$yr for $\alpha = 0.1$, $\beta = 1.0$, and $\gamma = 4$. At present time, there is only one unstable collapsing LMC, while at the end of simulation, there are five unstable LMCs. Thus, the simulation shows that the core D of TMC-1 will be a cluster of five gravitationally unstable protostellar cores with masses $\sim 3 \times 0.8 + 2 \times 0.7 = 3.8 M_\odot$.

## 4. Summary and conclusion

Star formation is a great theme in astronomy today, and special attention is attributed to the formation of low-mass stellar groups. Large ground-based telescopes and currently active satellite observatories have revealed many low-mass condensations (LMCs) in the molecular cores. It seems that the collisions and merger of these LMCs may lead to formation of moderate mass objects and eventually to formation of small groups of stars within the parent core. Here, we made a try to investigate the future of these LMCs. For this purpose, as an excellent test-place, we applied the observed substructure fragments in the core D of TMC-1, which their radii and masses at the present time are shown in Fig. 1.

Initial positions of LMCs in the cylindrical parent core are depicted in Fig. 2, in which their initial axial distances are randomly chosen between $0.4R_0$ and $0.7R_0$ where $R_0$ is the radius of cylinder. The observations show that the LMCs have infall velocities in the sub-sonic regime. Here, we assumed that there is two free parameters for the velocity vector of LMCs: transverse component parameter $\alpha$ and axial inclination parameter $\beta$. The radial and transverse components of the initial velocity vector of each LMC are schematically shown in Fig. 3. The acceleration (gravitational attraction and drag force) of LMCs are represented in terms of the fourth-order polynomial, and we used a merger parameter $\gamma$ to state the coalescence of them.

There are three key parameters $\alpha$, $\beta$, and $\gamma$ which control the properties of a model in the computer experiments. The results of computer experiments with a wide variety of these parameters show that overall the LMCs merger to form more massive larger bodies. As a direct result of merge, the number of LMCs decreases by time, and their mean mass increases as shown in Fig. 4 for a typical model. As another result of merger, the total mechanical energy of a system of LMCs increases by time until it approach to zero at a critical time $t_c$. This critical time increases by facilitating the merger of LMCs which arose by decreasing of the parameters $\alpha$, $\beta$, and $\gamma$ (Fig. 5).

At last, the mass distribution of LMCs are evaluated in the simulation time, and the results of a power law fit and its goodness-of-fit, with 50% error in the number of mass points, are shown in Fig. 6. On the whole, the results show that a system of LMCs in a low-mass protostellar core merge to form small groups of star-forming regions. The simulations show that the core D of TMC-1 turns about 40% of its mass into cluster of dynamically unstable protostellar cores. In general, we suggest that the future of LMCs in a core of molecular cloud is merger to covert about half of its initial masses into a cluster of gravitationally unstable protostellar cores. Clearly, over the collapse of each unstable protostellar core in the cluster, the mass conversion drops to few percent, consistent with the overall star formation efficiency of molecular clouds in the Galaxy. Although, the results of this research are deduced from simulation of LMCs in the core D of the TMC-1, but can reliably be applicable for future of all molecular cores with a system of LMCs.

## Acknowledgments

This work has been supported by Research Institute for Astronomy and Astrophysics of Maragha (RIAAM).

## A. Radius-mass relation for LMCs

We suppose that LMCs are pressure confined, self-gravitating isothermal gas spheres in hydrostatic equilibrium, which conform the Lane-Emden equation,

$$\frac{1}{\xi^2}\frac{d}{d\xi}\left(\xi^2\frac{d\psi}{d\xi}\right) = \exp(-\psi) \tag{A1}$$

where the dependent variable $\psi \equiv \phi/c_s^2$ is defined as the ratio of gravitational potential to the isothermal sound speed, and $\xi \equiv (4\pi G\rho_c/c_s^2)^{1/2}r$ is the nondimensional length where $\rho_c$ is the central density at $r = 0$. The equation (A1) can be solved numerically by the boundary conditions $\psi(0) = \psi'(0) = 0$, then the density profile can be obtained by $\rho = \rho_c \exp(-\psi)$. Since we assume that each LMC is confined by an external pressure $p_{ext}$, its edge with radius $\xi_{max}$ can be determined by the condition $\rho(\xi_{max}) = p_{ext}/c_s^2$. In this way, the nondimensional mass $m \equiv (p_{ext}^{1/2} G^{3/2}/c_s^4) M$ of each LMC is given by

$$m = \left(4\pi \frac{\rho_c}{\rho(\xi_{max})}\right) \left(\xi^2 \frac{d\psi}{d\xi}\right)_{\xi=\xi_{max}}. \tag{A2}$$

The Fig. 8 shows the density contrast $\rho_c/\rho(\xi_{max})$ and radius of pressure-bounded isothermal spheres as a function of nondimensional mass. Masses smaller than the critical value $m = 1.18$ are stable while ones with $\xi_{max} > 6.5$ (or equivalently $\rho_c/\rho(\xi_{max}) > 14.1$) are gravitationally unstable so that may collapse. Here, we approximately use the linear fitted relations to express the numerical results of stable states of Fig. 8 (in the range of $0 < m < 1.18$) as follows:

$$\xi_{max} = 5.51m, \quad \log_{10}\left(\frac{\rho_c}{\rho(\xi_{max})}\right) = 0.97m. \tag{A3}$$

There is a variety range of densities reported in the TMC-1. For instance, around the cyanopolyyne peak, Schloerb et al. (1983) and Bujarrabal et al. (1981) found molecular hydrogen densities of $5-10\times10^4$ and $1-3\times10^4\mathrm{cm}^{-3}$, respectively, both from multi-transition studies of $HC_3N$. Pratap et al. (1997) found that over most of the TMC-1 ridge the density was relatively uniform with a value of approximately $6 \times 10^4\mathrm{cm}^{-3}$. PLVKL used the CCS column densities to distinguish LMCs with an $H_2$ densities of $\sim$ a few $\times10^4\mathrm{cm}^{-3}$. Here, we apply an appropriate mean value of $\sim 5 \times 10^4\mathrm{cm}^{-3}$ for the ambient gas of the TMC-1, thus we have

$$R = 1.7 \left(\frac{T}{10\mathrm{K}}\right) \left(\frac{n_c}{5 \times 10^4\mathrm{cm}^{-3}}\right)^{-1/2} \xi_{max} \tag{A4}$$

and

$$M = 0.5 \left(\frac{T}{10\mathrm{K}}\right)^{3/2} \left(\frac{n(\xi_{max})}{5 \times 10^4\mathrm{cm}^{-3}}\right)^{-1/2} m \tag{A5}$$

for radius and mass of each LMC in units of $u_l$ and $u_m$, respectively. Eliminating the $m$ between (A4) and (A5) by using the linear approximations (A3), and choosing $T = 10\mathrm{K}$ and $n(\xi_{max}) = 5 \times 10^4\mathrm{cm}^{-3}$ for TMC-1, we approximately have

$$R \approx 18.7M10^{-M}. \tag{A6}$$

In this way, the dimensional radius can directly be obtained via equation (A6) for merged stable LMCs with total mass $M_i + M_j < 0.5 \times 1.18$. Thus, not only the dimensional radius of the merged LMCs are obtained, but also its condition for collapse can be checked.

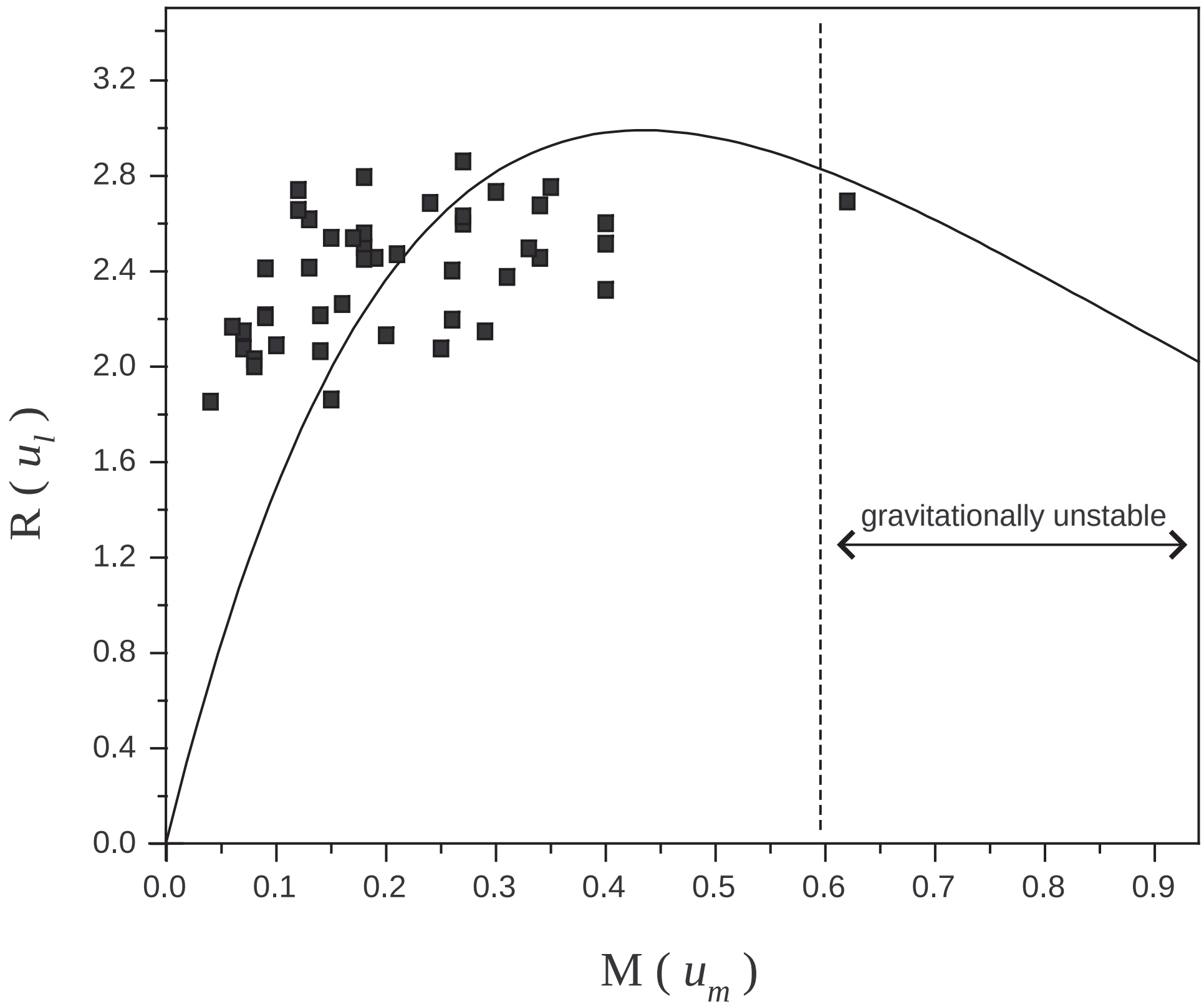

Fig. 1.— The radii of the 45 LMCs in the core D of TMC-1 and their masses at the present time $t = 0$ are shown as the black squares. The curve is the function $R = 18.7M10^{-M}$ which is explained in the Appendix A.

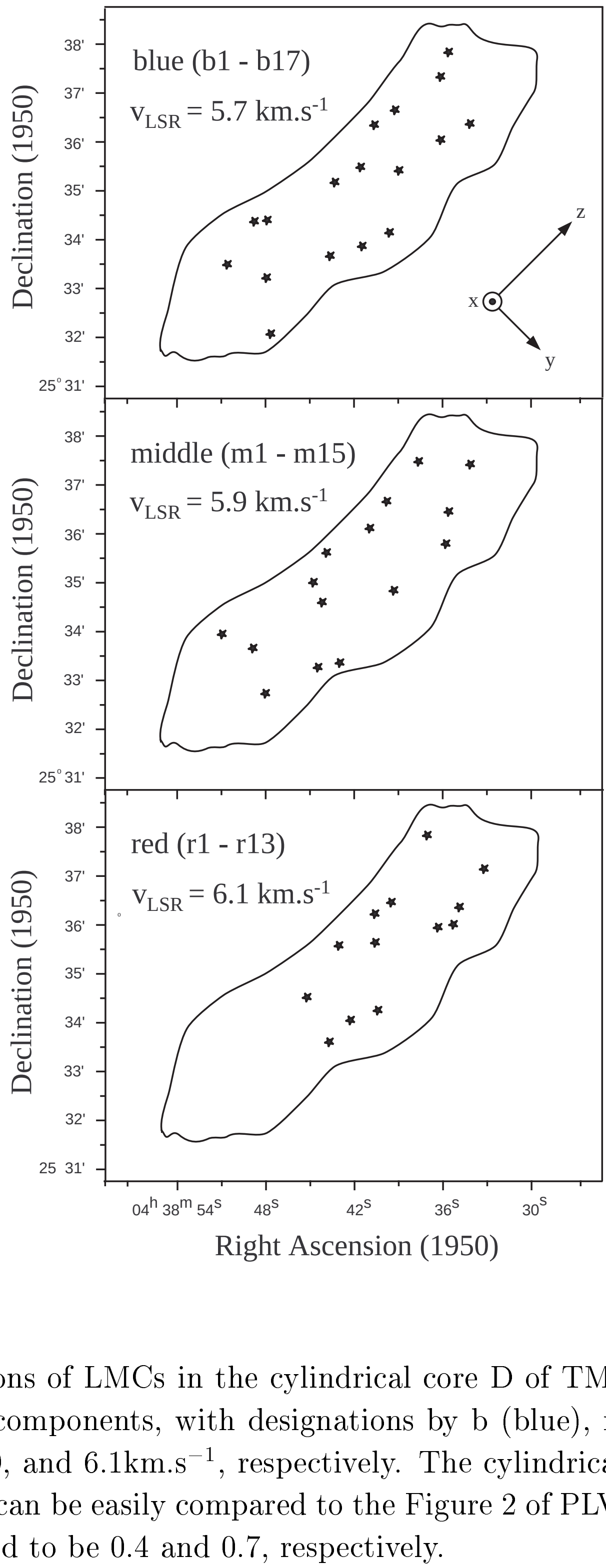

Fig. 2.— Initial positions of LMCs in the cylindrical core D of TMC-1. They are grouped by their LSR velocity components, with designations by b (blue), m (middle), and r (red) referring to the 5.7, 5.9, and 6.1km.s$^{-1}$, respectively. The cylindrical parent core is rotated in this figure so that it can be easily compared to the Figure 2 of PLVKL. Here, the fractions $f_{in}$ and $f_{out}$ are assumed to be 0.4 and 0.7, respectively.

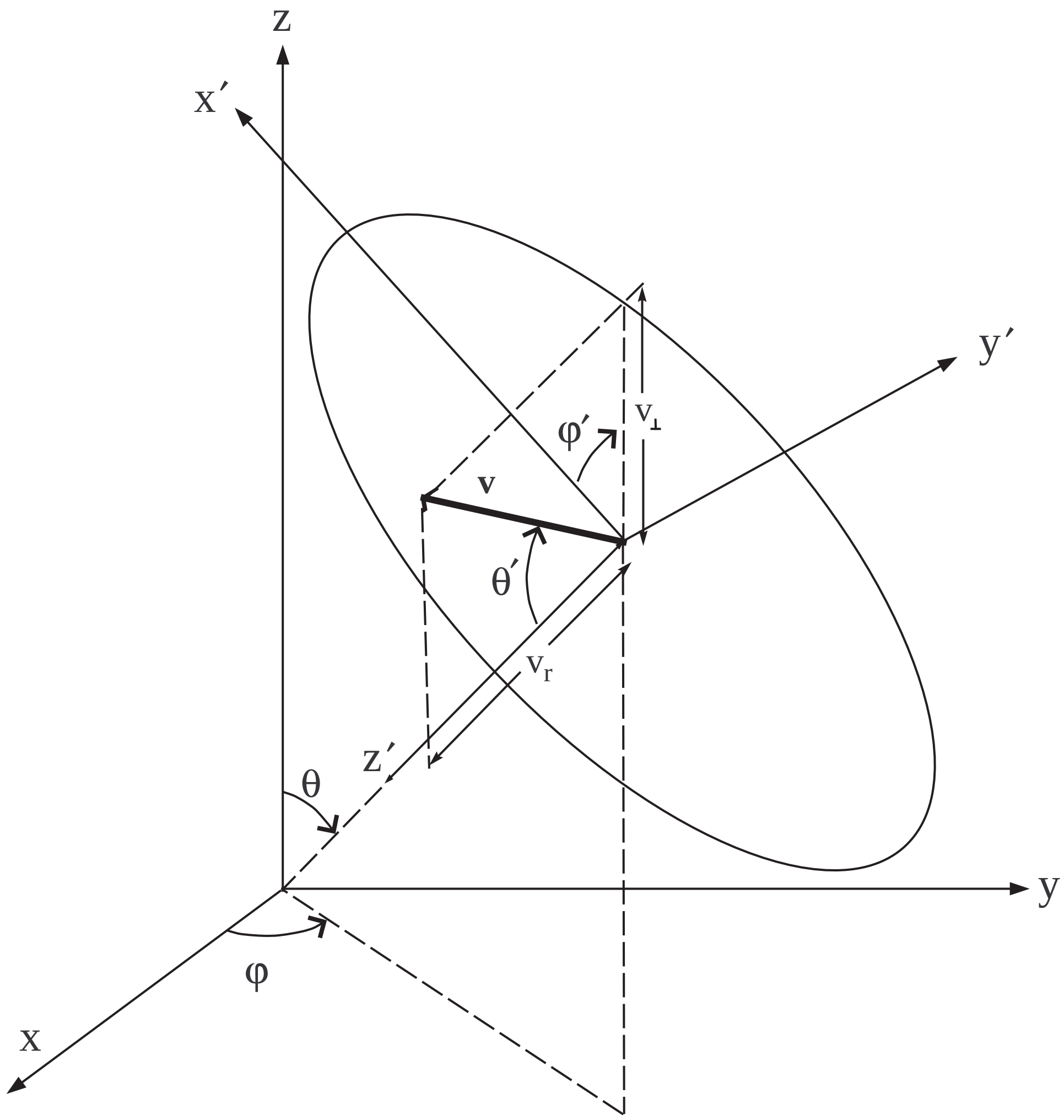

Fig. 3.— Schematic diagram of the radial and transverse components of the initial velocity vector of each LMC. The origin of $x'y'z'$-coordinate is placed on the LMC, and the origin of $xyz$-coordinate is assumed to be at the center of the core.

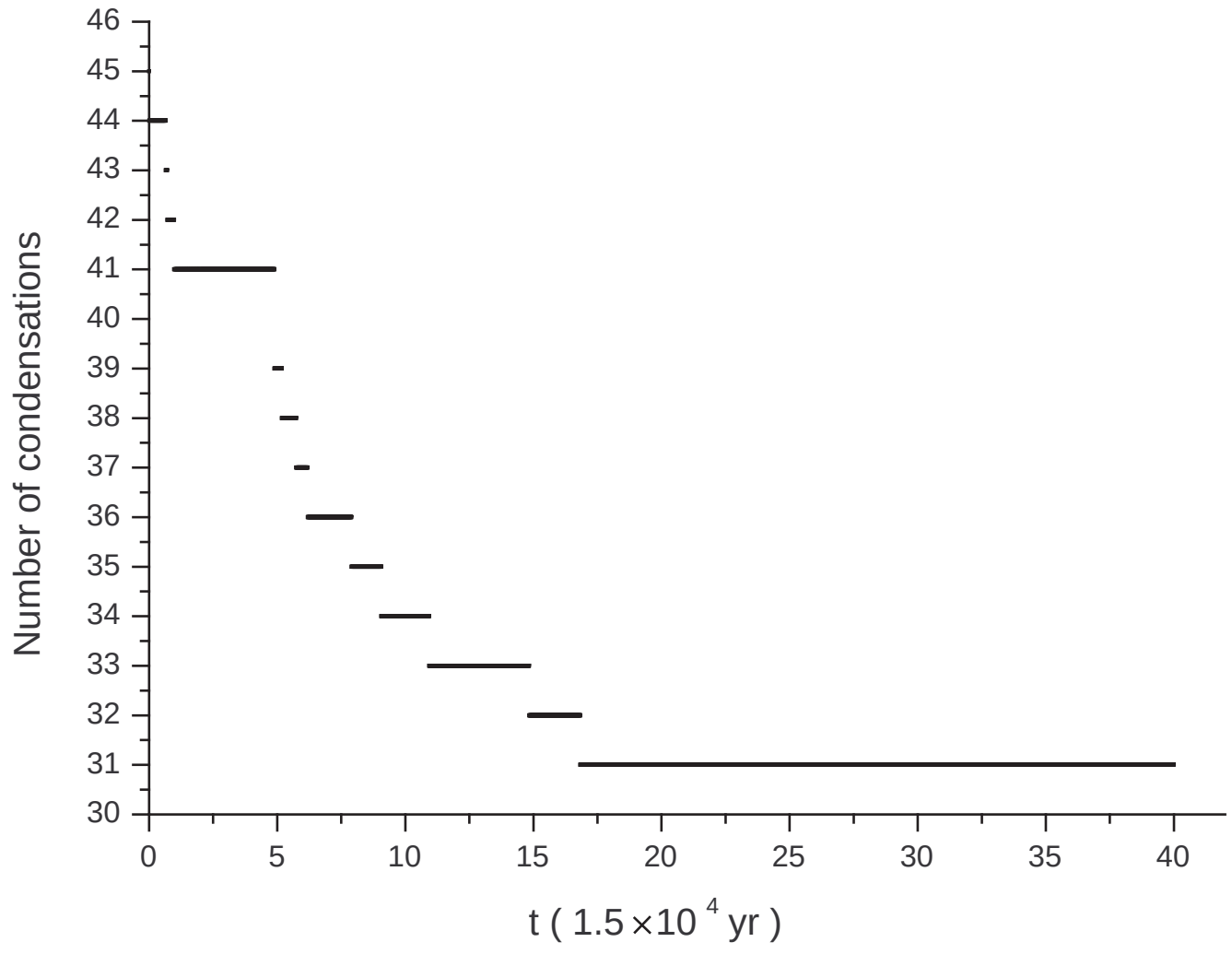

(a)

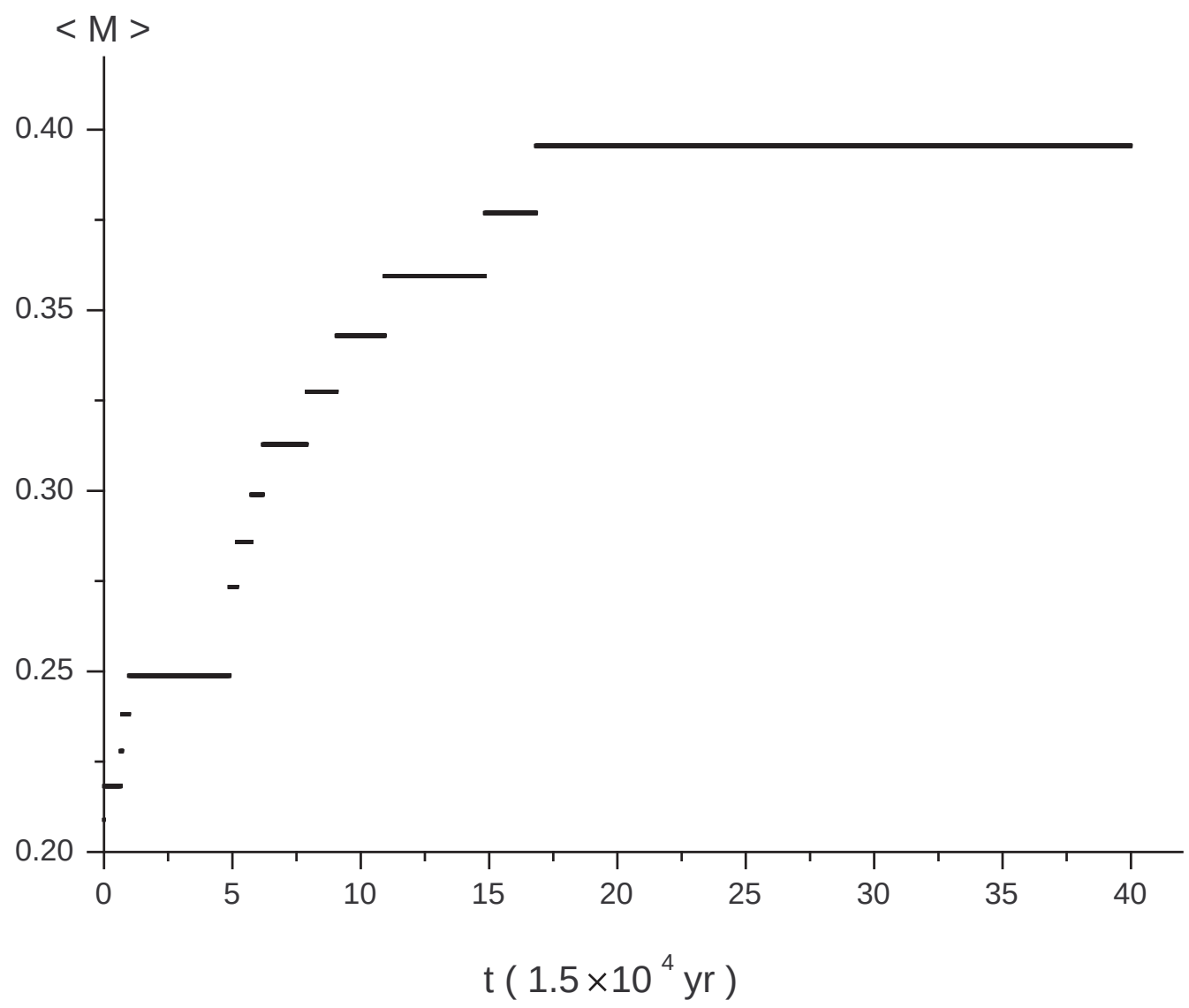

(b)

Fig. 4.— (a) Number of LMCs, and (b) the mean mass of them versus the simulation time, for $\alpha = 0.1$, $\beta = 1.0$, and $\gamma = 4$.

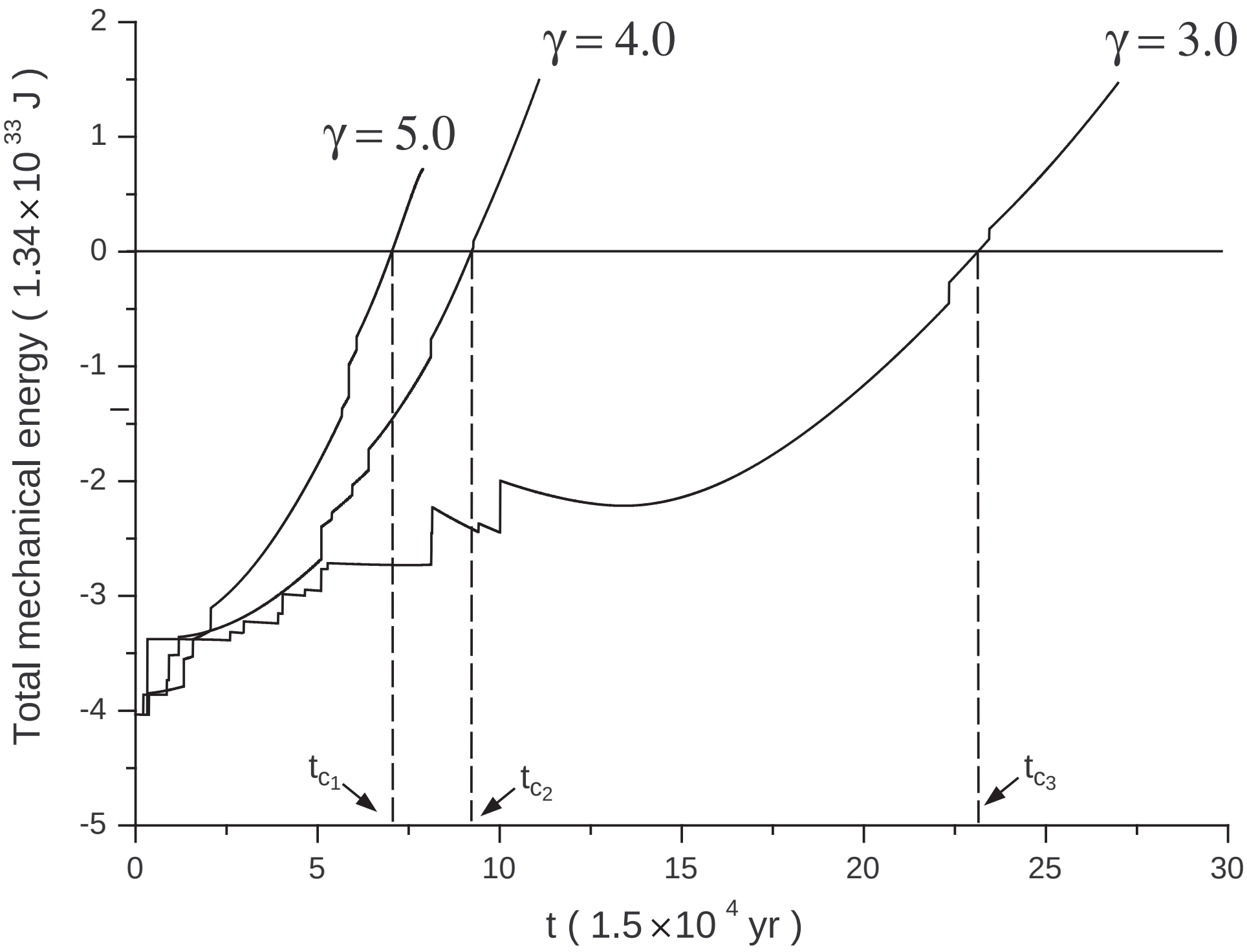

Fig. 5.— Total mechanical energy of LMCs versus simulation time for three typical values of merger parameter $\gamma$. The parameters $\alpha$ and $\beta$ are 0.1 and 1.0, respectively. Clearly, decreasing the parameter $\gamma$ facilitates the merger of LMCs.

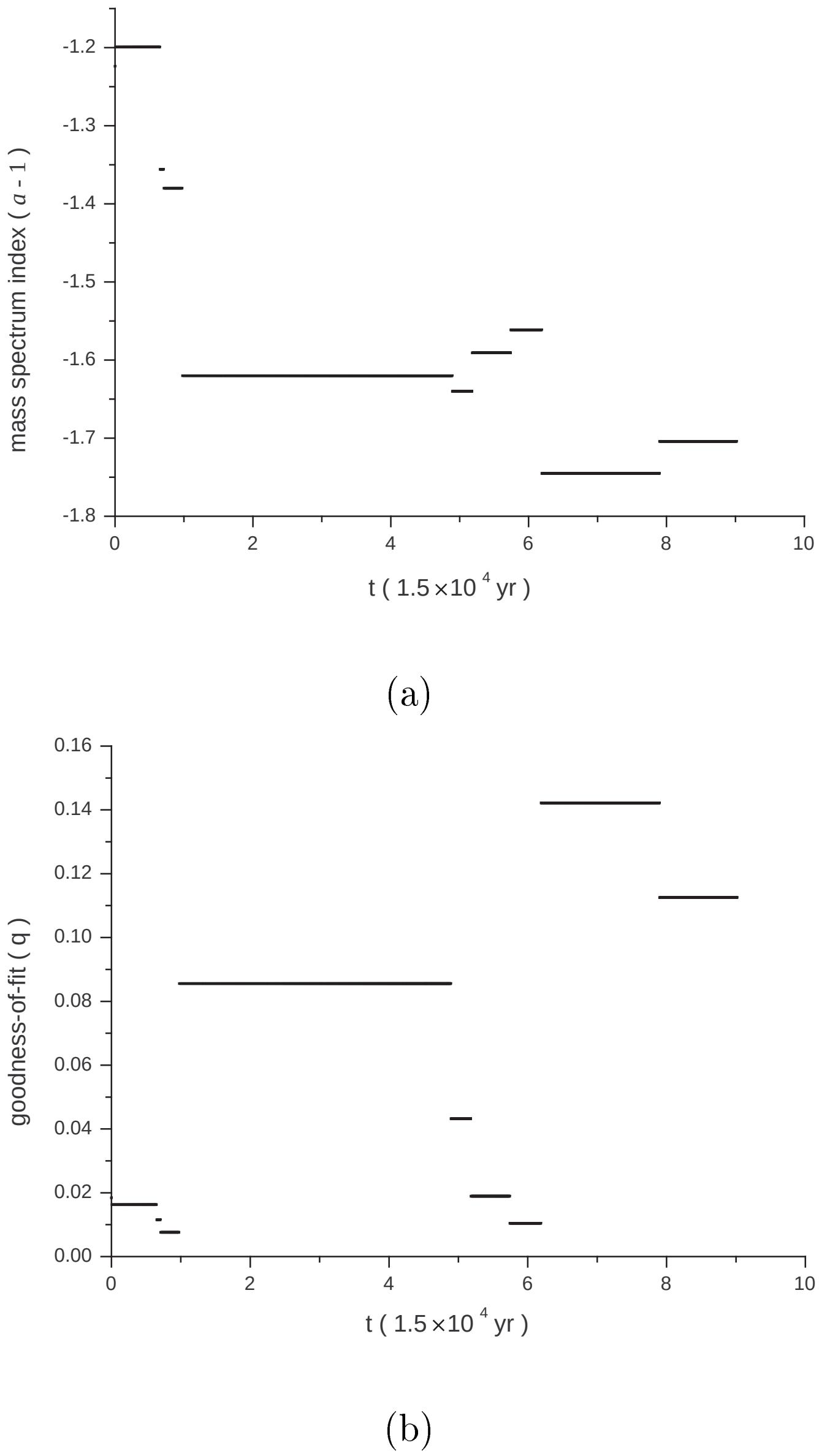

Fig. 6.— (a) Mass spectrum index versus simulation time with $\gamma = 4$, $\alpha = 0.1$ and $\beta = 1.0$. (b) The goodness-of-fit with assumption of 50% error in the number of LMCs.

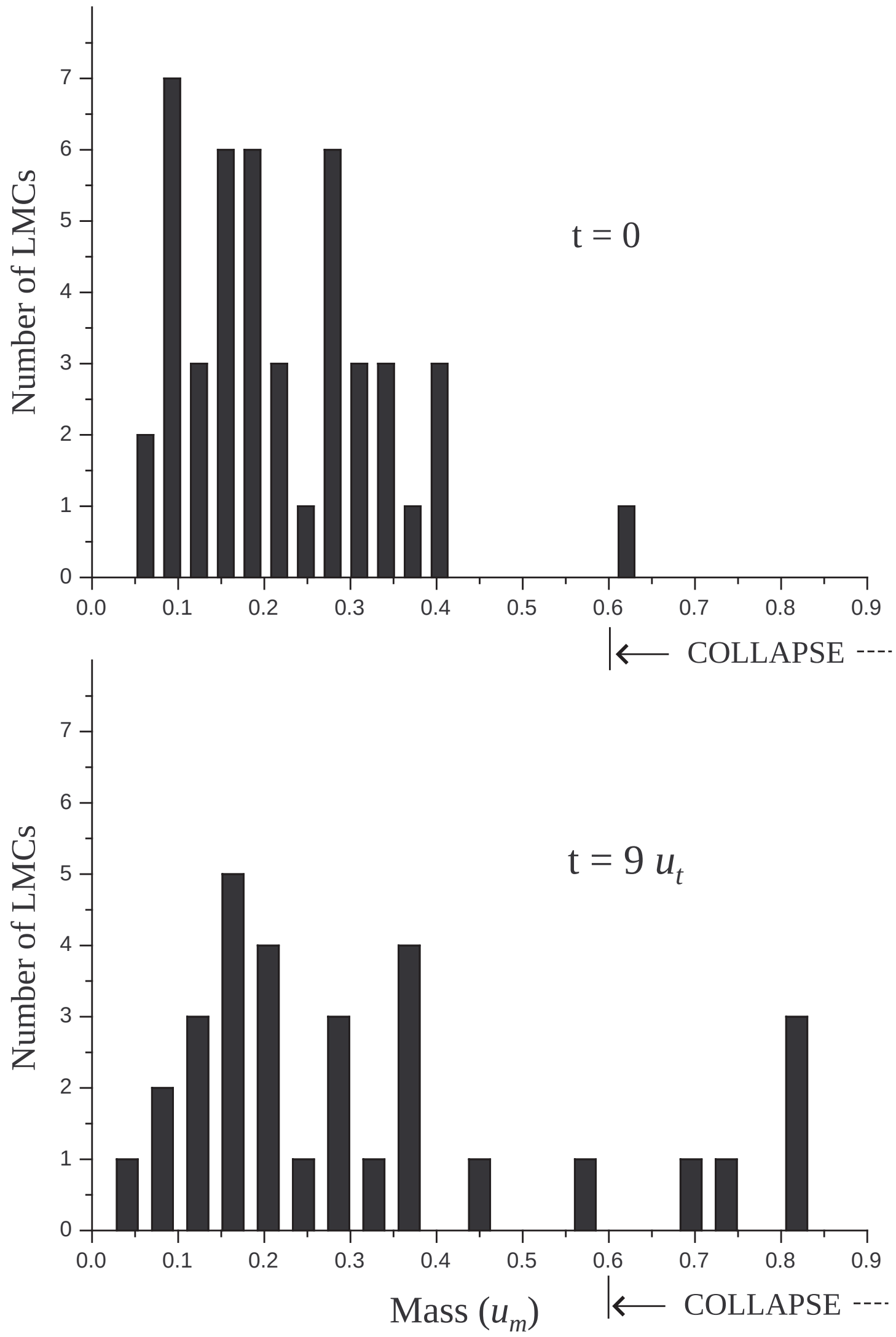

Fig. 7.— Histogram of the mass distribution of LMCs at present time (top panel), and at the simulation time $t \approx 1.5 \times 10^5$yr (bottom panel) for $\alpha = 0.1$, $\beta = 1.0$, and $\gamma = 4$.

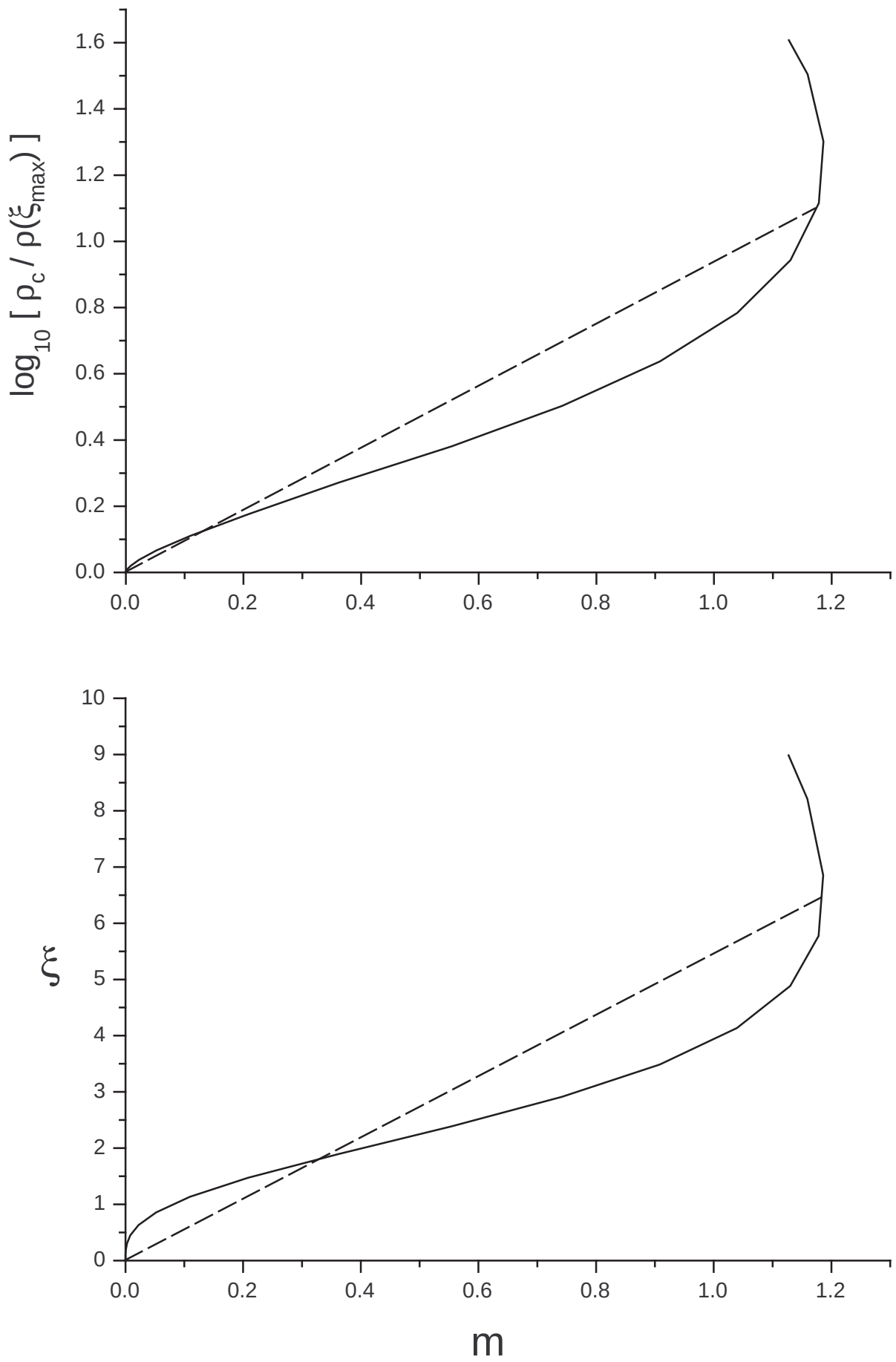

Fig. 8.— The density contrast $\rho_c/\rho(\xi_{max})$ and the nondimensional radius of pressure-bounded isothermal spheres as a function of nondimensional mass. The dashed lines are approximately linear relations to the numerical results for the gravitationally stable cases.